\documentstyle[preprint,prd,aps,amsfonts,psfig]{revtex}
\begin{document}

\title{
\vspace*{0.2in}
Multiplicity distributions at high energies as a sum of
 Poissonian-like distributions}
\draft
\author{O.~G. Tchikilev}

\address{Institute for High Energy Physics, Moscow region,
142284 Protvino, Russia}

\date{\today}

\maketitle

\begin{abstract}
It is shown that at collider energies  experimental distributions in the
multiplicity $n$ of negatively charged particles in inelastic and
non-single diffractive $\overline{p}p$ collisions are well parameterized 
by a sum of so-called Gupta-Sarma distributions having the Poisson 
distribution as a particular case.
 This extends the  earlier description  of the multiplicity distributions
in hadron-hadron collisions at c.m. energies below 65~GeV
by the two parameter sum of Poissonians. Implications of the proposed
parametrization for the LHC energy are discussed.
\end{abstract}


\section {Introduction}

 Charged particle multiplicity distributions in inelastic
 hadron-hadron collisions at high energies are usually described by
 the negative binomial (NB), for an experimental review
 see Ref. \cite{geich}. However the appearance of the shoulder
 structure observed for the first time by the UA5 Collaboration \cite{UA5}
 has led to the  use of the weighted sum of two NB's  with
 five free parameters \cite{fugles}, where the first NB describes the
 contribution of soft events (events without
 minijets) and  the second one describes the contribution
 of semihard events (events 
 with minijets). 
The aim of this paper is to extend to the collider energies another
 phenomenological parametrization giving better agreement with lower 
energy data than NB both for $p(\overline{p})p$\cite{chliap1,chliap2}
and meson-proton\cite{chliap3,chliap2} collisions. In the 
Refs.~\cite{chliap1,chliap2} the multiplicity distribution $P_{n}$
of negatively charged particles produced in inelastic $p(\overline{p})p$
collisions at the center of mass energies $\sqrt{s}$ below 63~GeV have
been fairly well described by a two parameter sum of Poissonians. This
approach is based on a simple minded quark-parton model in
  which quarks~$q$ interact pairwise independently of one another 
 with the same conditional probability $\varepsilon$ and each
 $qq$ interaction leads to the same multiplicity distribution in the 
 final state. The probabilities
 for events with 0, 1, 2 or 3 $qq$ interactions are equal respectively to
 $(1 - \varepsilon )^{3}$, $3 \varepsilon (1 - \varepsilon ) ^{2}$,
 $3 \varepsilon ^{2} (1 - \varepsilon )$ and $\varepsilon ^{3}$ and in
 terms of a probability generating function, p.g.f. (for a mathematical
 formalism see Refs.~\cite{carru,dremin}) it leads to the relation

\begin{equation}
 G(z) = \sum{P_{n} z^{n}} = (1- \varepsilon + \varepsilon \varphi (z))^3~,
\end{equation}

\noindent
where $G(z)$ is the p.g.f. for the final distributon and 
$\varphi (z)$ is the p.g.f. for events with one parton-parton collision.
The p.g.f.'s for events with two or three parton-parton collisions are
simply convolutions $\varphi ^2 (z)$ and $\varphi ^3 (z)$. Good description
 of the experimental data has been obtained with $\varphi (z) = 
\exp {( S ( z - 1 ) )}$, the p.g.f. for the Poisson distribution.
This parametrization had the strong energy dependence of
 the parameter $\varepsilon$~\cite{chliap1}, more smooth energy 
dependence of the $\varepsilon$ was observed \cite{chliap2}
 when the Poissonian was replaced by the Poisson
distribution truncated at zero multiplicity with the p.g.f.
 $\varphi ' (z) = (\varphi (z) - \varphi (0) )/ ( 1 - \varphi (0) )$. 
The need for the truncation at zero is explained for $pp$ collisions by
non-zero electric charge of the initial $uu$, $ud$ and $dd$ pairs leading
to the reduced probability to have zero charged particles in the final
state. This is not the case for $\overline{p}p$ collisions where
$\overline{u}u$ and $\overline{d}d$ pairs (5 out of 9 combinations) are
neutral.

 The model~\cite{chliap1,chliap2,chliap3}
 is modified in the present paper in the two aspects:

 a) The Poisson distribution is replaced by the so-called Gupta-Sarma
 (GS) distribution \cite{sarma1,sarma2} with the p.d.f.

\begin{equation}
 g(z)=\exp {\biggl({{  -S\, ( 1 - z )}\over{ 1 + r\,( 1 - z)}} \biggr)}~.
\end{equation}

\noindent
The GS distribution having the Poissonian as a particular case at $r=0$
and known in mathematical statistics under the name of the
 P\'olya-Aeppli distribution \cite{JKK93} has physical interpretation in the
framework of different models \cite{biya,finkel,chau1}, see discussion in
Refs.~\cite{SARMA3,HUANG}. In the approach advocated by Biyajima 
et al.~\cite{biya} and Finkelstein~\cite{finkel} the multiplicity
distribution originates from the Poisson distribution of some clusters, each
cluster obeys Furry-Yule (or truncated at zero Bose-Einstein) distribution,
finally it leads to the p.g.f. (2). One can note that the same form (2)
is valid when the p.d.f. for cluster decay distribution is a linear
fraction   $(1 + \Delta \, (1-z))/(1+ r \, (1-z))$, usual for the theory
 of branching processes, Ref.~\cite{bartlett}. 
In the Gupta-Sarma approach
\cite{sarma1,sarma2} the system after collision is viewed as one
highly excited hadron emitting entity obeying the branching process with
the probability per infinitesimal time $\Delta t$ to produce $k$ new particles
proportional to $\lambda ^k \Delta t$, where $\lambda $ is positive constant.
The solution of the corresponding evolution equation for initial condition
with zero particles leads to the p.g.f. of the form (2). In the Chau-Huang
approach \cite{chau1} the GS distribution is obtained from the statistical
Ising model.

 b) As suggested in the Ref.~\cite{chliap2}  events with zero parton-parton
collisions can represent the diffractive-like processes, with the fraction
of diffractive-like events given by the $( 1 - \varepsilon) ^3$.
It has been established that the multiplicity distribution for diffractive 
system with effective mass $M$ looks like the multiplicity distribution
in $pp$ collisions at the c.m. energy $\sqrt {s} = M$, see Ref.~\cite{goul}
for a review on diffraction. In this paper we approximate the diffractive
contribution by the form (1) with $\varphi (z)$ equal to the p.g.f. for the 
Poissonian under the crude assumption that the integrated over $M$ 
distribution is similar to the distribution at some effective mass 
$\overline{M}$
 and  the final p.g.f. is given by 

\begin{equation}
 G(z) = (1-\varepsilon)^3\, \varphi_d + 3\varepsilon (1-\varepsilon)^2\, g(z)
 + 3\varepsilon^2 (1-\varepsilon)\, g^2(z) + \varepsilon^3\, g^3(z)
\end{equation}

\noindent
with 

\begin{equation}
 \varphi_d (z) = (1-\varepsilon + \varepsilon\, \exp (\, S_d\, (z-1)))^3.
\end{equation}

\noindent
 More careful description of the diffractive-like events
 with integration over $M$ is given in Refs. \cite{cool,ham}.

 In the section II the main characteristics of the Gupta-Sarma distribution
 are summarized.
 In the section III we present results of fits to the available $pp$ data
 and to the $\overline{p}p$ data obtained
 at the $S\overline{p}pS$ collider. In the section IV the discussion and
 the conclusions are given.

\section{Characteristics of the Gupta-Sarma distribution.}

 The mean multiplicity $<n>$ and its dispersion $D = (<n^2>-<n>^2)^{1/2}$
 are easily obtained from the p.d.g. (2) using formulae 
 $<n> = d\,g/d\,z \mid _{z=1} $ and
 $ D^2 = <n> + d^2 \ln g(z)/d\,z^2 \mid _{z=1}$

\begin{equation}
  <n> = S, ~~~~~~~~~~~ D^2 = <n> \, (1+ 2\, r).
\end{equation}

\noindent 
 To obtain expressions for probabilites $g_{n}$ one can use the method
proposed by Finkelstein\cite{finkel}. The p.g.f. is expressed as a sum
of powers $(z/(1+r\,(1-z)))^k$, where the denominator represents the well
known NB. Then, the contribution to $g_{n}$ 
with $n \ne 0$ from the $k$-th term 
is equal to the NB probability to have $n-k$ particles and finally it
gives

\begin{equation}  
 g_n =g_0\, \sum_{k=1}^{n} \frac{(n-1)!}{(n-k)!\,(k-1)\,!k!\,}\, a^k \, b^{n-k}
\end{equation}

\noindent
 with

\begin{equation}
  a= \frac{S}{(1+r)^2}~, ~~~~~   b = \frac{r}{1+r}~,
\end{equation}

\noindent
and
\begin{equation}
   g_0 = g(0) = \exp {\biggl( \frac{-S}{1+r} \biggr)}~.
\end{equation}

\noindent

 One can note \cite{biya,SARMA3} that the GS distribution is a particular
case of the partially coherent laser distribution PCLD (see review of the
PCLD in \cite{carru}). Indeed the p.g.f. for the PCLD is the product of the
p.d.f. (2) and $(1+ r \, (1-z))^{-k}$, i.e. the convolution of the GS and
NB distributions. It gives the expressions for $g_n$ in terms of the
Laguerre polynomials\cite{bateman}

\begin{equation}
g_n={\biggl(\frac{r}{1+r}\biggr)}^n\,\exp {\biggl(\frac{-S}{1+r} \biggr)}\,
L^{-1}_{n}{\biggl(\frac{-S}{r\,(1+r)}\biggr)}~.
\end{equation}

\noindent
 Using iteration relations for the Laguerre polinomials \cite{bateman} one
can obtain next iteration relations for $g_n$

\begin{equation}
  (n+1)\, g_{n+1} = (a\, + 2n\, b)\, g_n - (n-1)\, b^2\, g_{n-1}
\end{equation}
 
\noindent
 at $n > 1$ and

\begin{equation}
  g_1 =a\, g_0.
\end{equation}

\noindent
These iteration relations can be useful for calculations at large $n$ values.

\section{Results of fits}

 Both for $pp$ and $\overline{p}p$ data we calculate the number of
 negatively charged particles as $n=(n_{ch}-2)/2$, i.e. we count the number
 of produced pairs of charged particles. As mentioned above, we
 truncate the distribution for parton-parton collision in the fits to the
 $pp$ data and do not  truncate it for $\overline{p}p$ data 
 at the collider energies.

 The  $pp$ data used~[22-39] are the same as in Refs. \cite{chliap1,chliap2}
 with additional measurement from
 Ref.~\cite{17}.  In the Table 1 the results of fits
 to the distribution with the p.g.f. (3) are given for the case $r=0$. The
 agreement with experimental data is good, this is expected since even
 the two-parametr parametrization with the diffractive contribution
 concentrated at zero multiplicity was successful~\cite{chliap1,chliap2}.
 One can note also that the mean multiplicity for diffractive contribution,
  proportional to $S_d$ increases slowly with energy.

   Non-single diffractive (NSD) multiplicity distributions, measured by
the UA5 Collaboration~\cite{UA5,ua51} have been parameterized by the
distribution (3) without diffractive component, i.e. by the distribution
with three other ``parton-parton collision'' components normalized by
the factor $(1- (1-\varepsilon)^3)^{-1}$. Inelastic multiplicity
 distribution at $\sqrt s = 546$~GeV~\cite{ua51} has been parameterized
by the full distribution (3). The results of the fits are given in the 
table~2 both for  $\varepsilon = 0.456$ and for free $\varepsilon$.
The fixed value of the $\varepsilon$ was chosen on the assumption that
the fraction of diffractive like events is equal to 16\%, the fraction
of the single diffractive events  measured by the CDF Collaboration
at 546~GeV\cite{cdf}. The fraction measured by the UA5 
Collaboration is equal to 11~\%~\cite{ua51}, corresponding 
conditional probability $\varepsilon$ is equal to 0.52. 

The results of the fits with free $\varepsilon$
are illustrated in Figs. 1-3 respectively
for  inelastic and NSD data at 546~GeV\cite{ua51} 
and for NSD data at 900~GeV\cite{UA5}. The quality of the fits is quite qood,
the fluctuations in the parameter $\varepsilon$ are explained by
the our crude treatment of the diffractive component and possible bias in the
experimental data. We have ignored also the nonnegligible
contribution of the double diffraction processes in the NSD data.
 The influence of the high multiplicity
tail on the fit parameters has been observed also,
 the fit in the region $n_{ch} < 80$ of the inelastic 
data at 546~GeV gives more reasonable value of $S_d$ near 3, 
significantly smaller
than the values $S_d$ in the Table~2.

\section{Discussion and conclusions}

 The possibility that multiplicity distributions at high energies can split
 into several structures has been predicted more the 25 years ago\cite{abra}.
 For example Nielsen and Olesen made the statement\cite{nielsen}: ``if
 we go to high enough energy one should see a separation of the multiplicity
 spectrum in a series of equidistant peaks at $n \sim n_1$, $2n_1$, $3n_1$,
 ...''. Kaidalov and Ter-Martirosyan in the framework of the quark-gluon
 string model have predicted three peaks in the multiplicity
 distribution at $\sqrt {s}=100$~TeV~\cite{kaida}. These Regge-type 
 models \cite{abra,kaida} in principle predict more than three 
 peaks (structures) in contrast to our approach with maximum three
 nondiffractive structures. The prediction for NSD multiplicity
 distribution at LHC energy $\sqrt s = 14$~TeV, calculated with
 parameters $\varepsilon =0.456$, $r=0.8$ and $S$ fixed by the expected
 mean multiplicity $<n_{ch}> = 67.2$ is given in the Fig.~4. It seems to
 be intermediate between the drastic predictions of the Regge-type
 models and more flat predictions based on the parametrization by
 the weighted sum of the two NB's~\cite{giov,felix}.

 In conclusion, the multiplicity distributions at collider energies
 have been fairly well parameterized by the sum of Poisson-like
 GS distributions, with one GS distribution describing the multiplicity
 distribution for one ``parton-parton'' collision. An attempt has been
 made  to connect the fractions of events with 1, 2 or 3 ``parton-parton''
 collisions with the fraction of diffractive-like events in a framework
 of the simple minded parton model.

\newpage
\newcommand{\qq}{\mbox{$Q^{2}$}}
\newcommand{\en}{\mbox{$\sqrt s$}}
\newcommand{\np}{\mbox{$n_p$}}
\newcommand{\pnch}{\mbox{$P(n_{ch})$}}
\newcommand{\nch}{\mbox{$n_{ch}$}}
\newcommand{\avm}{\mbox{$<n_{-}>$}}
\newcommand{\avn}{\mbox{$<n_{-}>$(fit)}}
\newcommand{\axi}{\mbox{$\chi^2$/NDF}}
\newcommand{\dda}{\mbox{$\times 10^{-1}$}}
\newcommand{\ddb}{\mbox{$\times 10^{-2}$}}
\newcommand{\ddc}{\mbox{$\times 10^{-3}$}}
\newcommand{\ddd}{\mbox{$\times 10^{-4}$}}
\newcommand{\dde}{\mbox{$\times 10^{-5}$}}
\newcommand{\ddf}{\mbox{$\times 10^{-6}$}}

\renewcommand{\arraystretch}{1.01}
\begin{table}[bth]
\caption{Results of the fits to the negatively charged particle 
multiplicity distributions in inelastic $pp$ collisions [22-39].}
\begin{center}
\begin{tabular}{|c|c|c|c|c|c|}
Ref. & $\sqrt s$ (GeV) &$\varepsilon$ &$S$ &$S_d$ 
&\axi \\
\hline\hline
 \cite{8} & 6.84 & 0.336$\pm 0.002$ &0.203$\pm 0.006$ & 0.032$\pm 0.013$ & 
      21.1/5            \\
 \cite{9} & 7.87 & 0.381$\pm 0.004$ &0.325$\pm 0.011$ & 0 & 3.9/5  \\ 
 \cite{10}& 9.78 & 0.441$\pm 0.015$ &0.521$\pm 0.040$ & 0 & 7.5/6   \\ 
 \cite{11}&10.69 & 0.470$\pm 0.014$ &0.500$\pm 0.044$ & 0 & 2.8/5   \\ 
 \cite{12}&11.46&0.470 $\pm 0.008$ &0.671$\pm 0.019$ &0.013$\pm 0.040$&18.6/7 \\
 \cite{13}&13.76&0.472$\pm 0.009$ &0.940$\pm 0.028$ &0 &
 19.1/7 \\ 
 \cite{14}& 13.90&0.483$\pm 0.010$ &0.855$\pm 0.032$&0.003$\pm 0.041$&7.3/7 \\
 \cite{15} & 16.66 & 0.491$\pm 0.006$ &1.170$\pm 0.021$  &0.012$\pm 0.021$ &
 18.3/10  \\
 \cite{16}&18.17&0.523$\pm 0.031$ &1.224$\pm 0.094$&0&
 2.4/7 \\
\cite{17}&19.42 &0.551$\pm 0.016$&1.197$\pm 0.054$ &0.074$\pm 0.074$&9.0/8  \\
\cite{18} &19.66& 0.538$\pm 0.011$ &1.243$\pm 0.034$ &0.070 $\pm 0.071$&
 7.3/10\\ 
 \cite{19} &21.7&0.512$\pm 0.011$&1.460$\pm 0.040$&0.135$\pm 0.063$&14.3/11  \\
 \cite{20} &23.76&0.563$\pm 0.012$&1.486$\pm 0.039$ &0.112$\pm 0.066$&10.4/11\\
 \cite{21} &23.88&0.561$\pm 0.019$&1.601$\pm 0.074$ &0.246$\pm 0.148$&11.2/10\\ 
 \cite{22} &26.0&0.577$\pm 0.010$&1.638$\pm 0.034$ &0.102$\pm 0.054$&8.4/10\\ 
 \cite{14} &27.6 &0.542$\pm 0.013$& 1.720$\pm 0.059$ &0.096$\pm 0.057$&
17.8/13\\ 
 \cite{23}&27.6&0.555 $\pm0.016$ &1.565$\pm 0.077$ &0.027$\pm 0.068$&4.3/9 \\
\cite{24}&30.4&0.528$\pm 0.021$&2.000$\pm 0.077$ &0.300$\pm 0.157$&3.1/14\\
\cite{25}&38.8&0.576$\pm 0.008$&2.059$\pm 0.031$ &0.273$\pm 0.085$&8.3/13 \\ 
\cite{24}&44.5&0.538$\pm 0.021$&2.441$\pm 0.086$&0.438$\pm 0.167$& 4.9/16 \\ 
\cite{24}&52.6&0.549$\pm 0.016$&2.647$\pm 0.068$&0.381$\pm 0.142$& 12.2/18 \\ 
\cite{24}&62.2&0.552$\pm 0.017$&2.887$\pm 0.070$&0.426$\pm 0.168$&16.7/17 \\
\end{tabular}
\renewcommand{\arraystretch}{1.0}
\end{center}
\end{table}
\begin{table}[bth]
\caption{Results of the fits to the negatively charged particle 
multiplicity distributions in $\overline{p}p$ interactions [2,40].
 }
\begin{center}
\begin{tabular}{|c|c|c|c|c|c|c|}
events & $\sqrt s$ (GeV) &$\varepsilon$ &$S$ &$S_d$ & r 
&\axi \\
\hline\hline
 NSD & 200 & 0.456 &5.965$\pm 0.076$ & & 0.170$\pm 0.056$ &  19.7/29  \\
     & & 0.264$\pm 0.038$ &7.441$\pm0.304$ & & 0.297$\pm0.084$& 9.2/28 \\    
 NSD & 546 & 0.456 &8.429$\pm 0.053$& & 0.468$\pm0.030$ & 61.3/45 \\
     & & 0.352$\pm0.018$&9.453$\pm0.187$& & 0.558$\pm0.038$& 32.6/44 \\
inel.&546 &0.456 &7.410$\pm0.091$&11.071$\pm0.458$&0.832$\pm0.075$& 39.4/44\\
 &&0.536$\pm0.027$&6.743$\pm0.067$&11.811$\pm0.326$&0.794$\pm0.067$&29.9/43\\
 NSD & 900& 0.456&10.400$\pm0.100$& & 0.703$\pm0.065$& 77.0/52\\   
   & & 0.304$\pm0.029$& 12.213$\pm0.394$& & 0.823$\pm0.095$& 20.9/51 \\
\end{tabular}
\renewcommand{\arraystretch}{1.0}
\end{center}
\end{table}
\clearpage

\begin{figure}
\centerline{\psfig{file=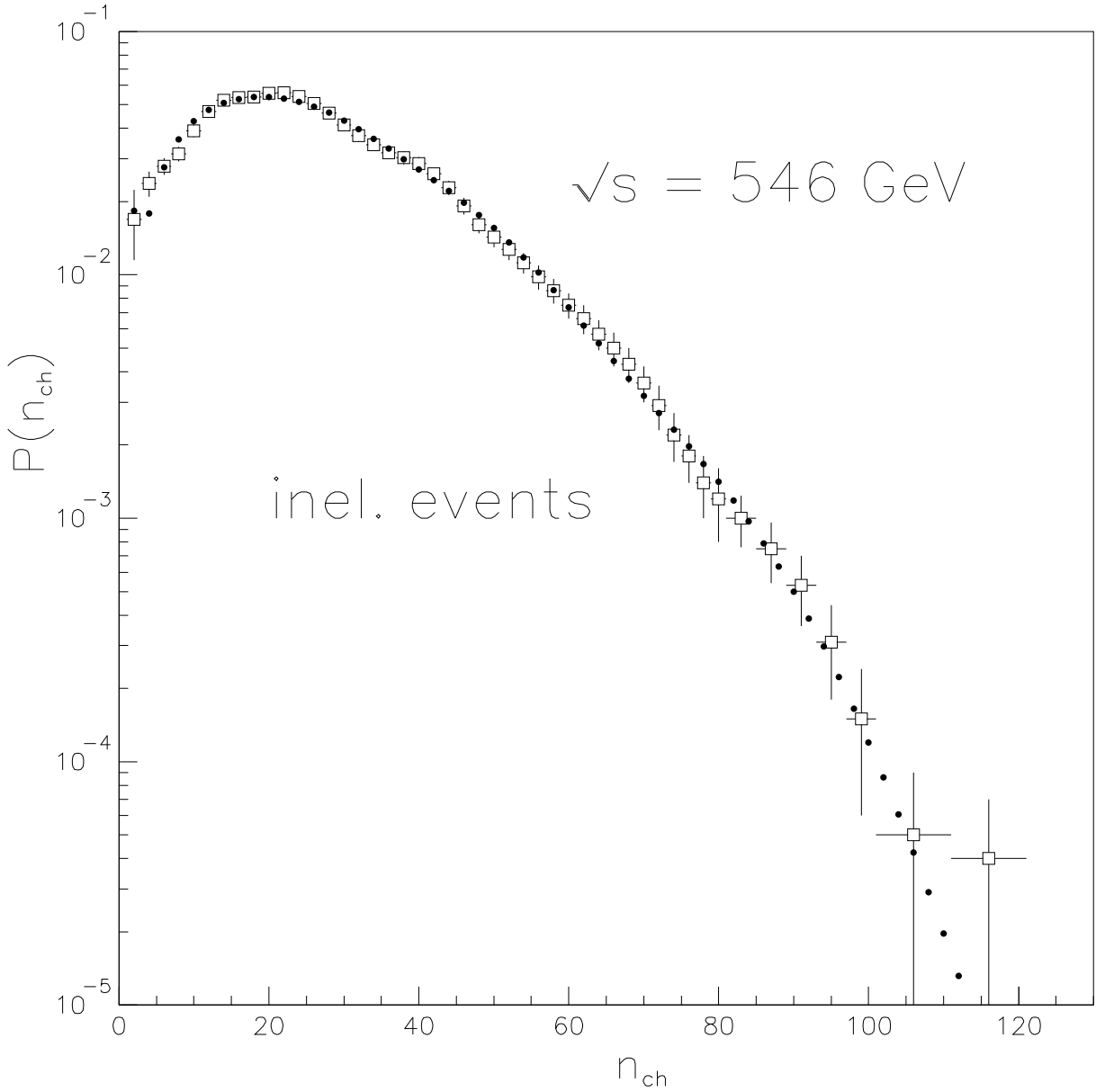,width=18cm}}
\caption{Multiplicity distribution for inelastic
 antiproton-proton collisions at the c.m. energy
  546~GeV~[40] (squares) compared with results of the
fit  (full dots).
}
\label{nee}
\end{figure}

\begin{figure}
\centerline{\psfig{file=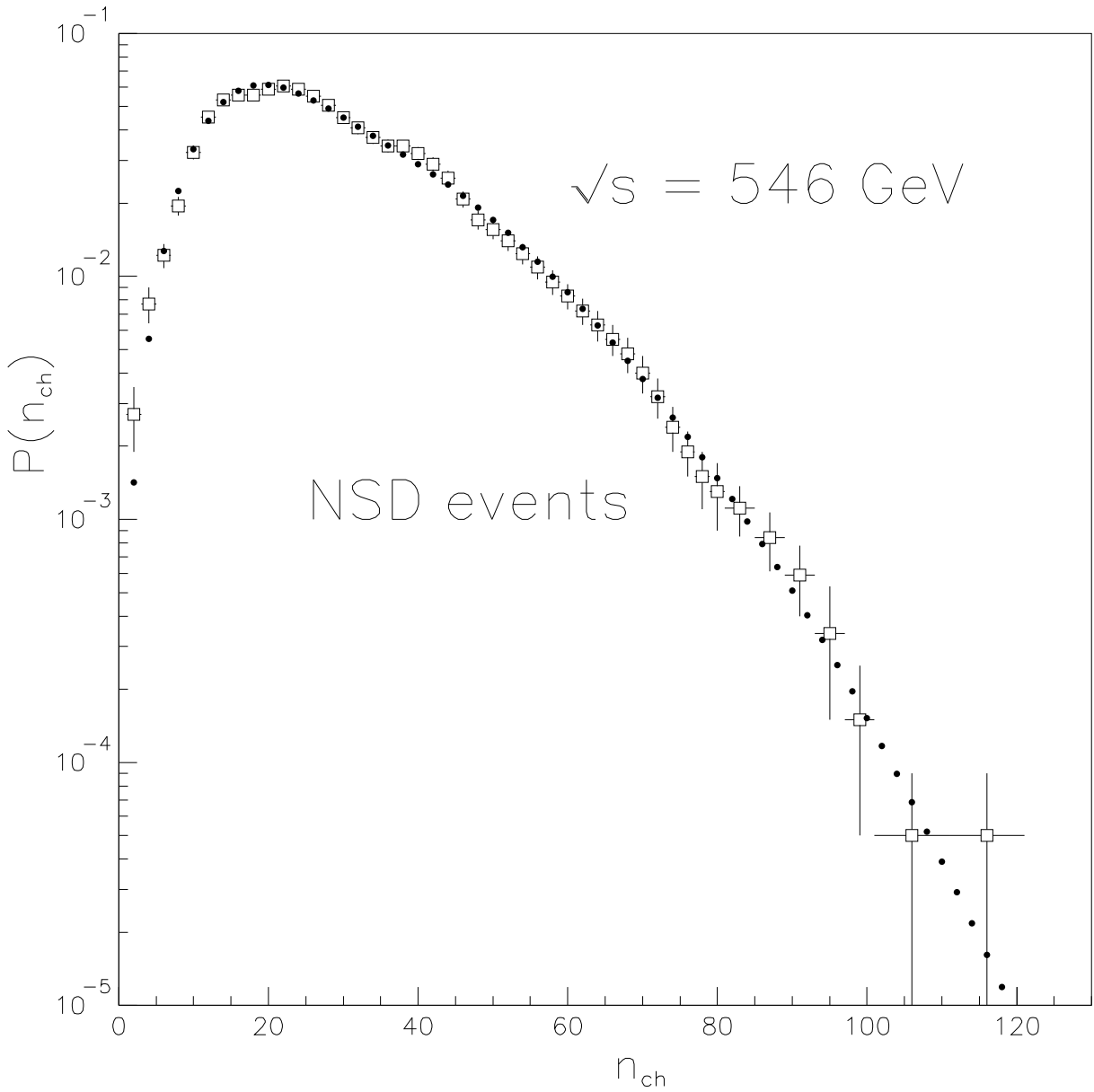,width=18cm}}
\caption{ Multiplicity distribution for NSD  collisions at the c.m. energy
  546~GeV~[40] (squares) compared with results of the fit (full dots).
}
\label{npp}
\end{figure}

\begin{figure}
\centerline{\psfig{file=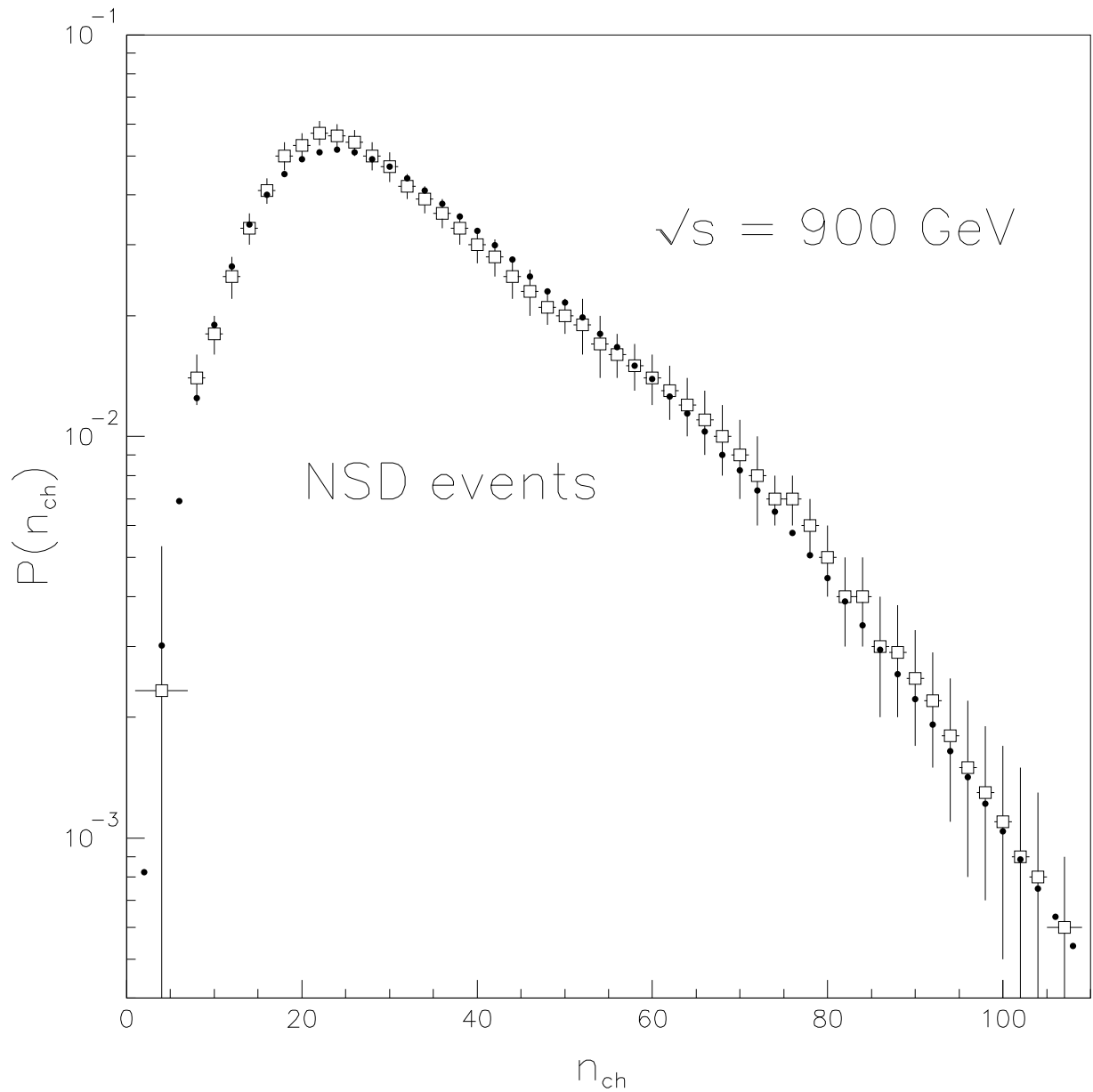,width=18cm}}
\caption{Charge particle
 multiplicity distribution  for NSD  collisions at the c.m. energy
 900~GeV~[2] (squares) compared with results of the
fit (full dots).
}
\label{cpp}
\end{figure}

\begin{figure}
\centerline{\psfig{file=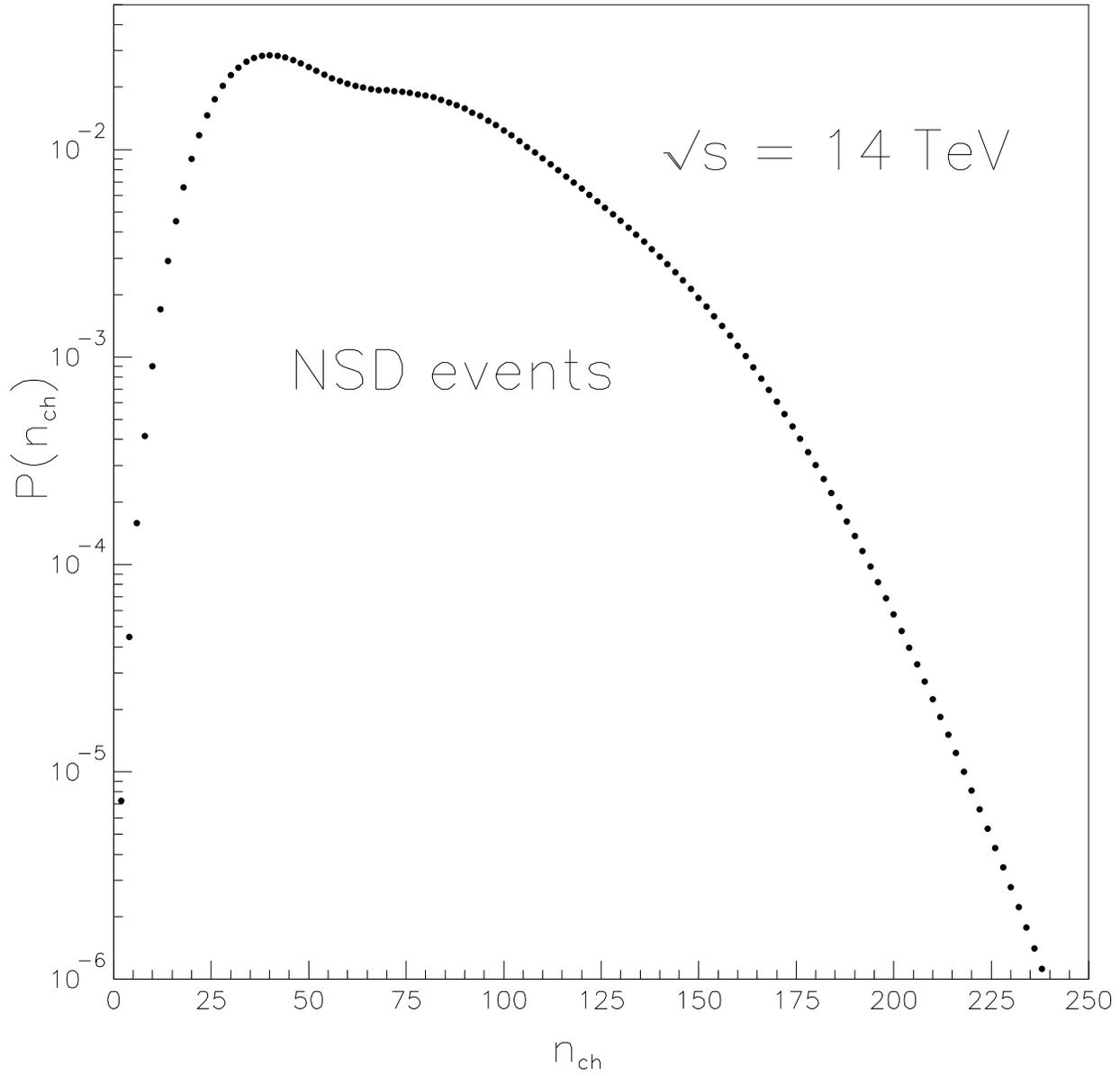,width=18cm}}
\caption{Prediction of the NSD charged particle multiplicity distribution
at the LHC energy  14~TeV.
}
\label{ptd}
\end{figure}

\end{document}